# Experiments and Modeling of the Autoignition of Methyl-Cyclohexane at High Pressure

B.W. Weber[1]   W.J. Pitz[2]   C.J. Sung[1]   M. Mehl[2]   E.J. Silke[2]   A.C. Davis[3]

[1]Department of Mechanical Engineering, University of Connecticut, Storrs, CT 06029
[2]Lawrence Livermore National Laboratory, Livermore, CA 94550
[3]Clean Combustion Research Center, King Abdullah University of Science and Technology, Thuwal, Makkah 23955-6900, Kingdom of Saudi Arabia

The autoignition delays of mixtures of methyl-cyclohexane (MCH), oxygen, nitrogen, and argon have been studied in a heated rapid compression machine under the conditions $P_C = 50$ bar, $T_C = 690 - 910$K. Three different mixture compositions were studied, with equivalence ratios ranging from $\phi = 0.5 - 1.5$. The trends of the ignition delay measured at 50 bar were similar to the trends measured in earlier experiments at $P_C = 15.1$ and 25.5 bar. The experimentally measured ignition delays were compared to a newly updated chemical kinetic model for the combustion of MCH. The model has been updated to include newly calculated reaction rates for much of the low-temperature chemistry. The agreement between the experiments and the model was substantially improved compared to a previous version of the model. Nevertheless, despite the encouraging improvements, work continues on further advances, e.g. in improving predictions of the first stage ignition delays.

1. **Introduction**

Cycloalkanes and alkyl-cycloalkanes are well known major components in several transportation fuels, including gasoline, diesel and jet fuels (Briker et al. 2001). Since these transportation fuels have hundreds or thousands of individual chemical components, incorporation of all these components in the kinetic model would make it very difficult to build and computationally expensive to use. To facilitate modeling such real fuels, it is necessary to formulate a surrogate mixture by selectively choosing a much smaller set of neat components that will reproduce the physical and chemical behavior of the target fuel. Methyl-cyclohexane (MCH) is frequently suggested as a candidate component in these formulations to represent the cycloalkane content of real fuels (Bieleveld et al. 2009; Naik et al. 2005). Furthermore, an understanding of MCH kinetics can provide the base from which to build models of the combustion of other cycloalkanes or alkyl-cycloalkanes.

If chemical kinetic models are to be used in engine design, it is critical that they are able to reproduce the combustion behavior of fuels under the thermodynamic conditions prevalent in engines. New engines, using advanced concepts such as homogeneous charge compression ignition (HCCI) and reactivity controlled compression ignition (RCCI), incorporate Low Temperature Combustion (LTC) to help achieve goals of improved fuel efficiency and lower emissions. However, a detailed understanding of LTC reaction pathways is often required to properly predict combustion phasing, heat release rates, and engine-out emissions. Also, HCCI and RCCI engines operate at high pressures and the chemical kinetic models used in engine simulations need to be validated at these conditions. Therefore, experimental data acquired at engine-relevant conditions are of critical importance for validating chemical kinetic model performance.

Previous work conducted in our Rapid Compression Machine (RCM) measured the ignition delays of MCH at pressures of 15.1 and 25.5 bar and for three equivalence ratios of $\phi = 0.5$, 1.0, and 1.5 (Mittal et al. 2009). Over the temperature range of 680–840 K, substantial discrepancies between the experimental data and the earlier Lawrence Livermore National Laboratories (LLNL) kinetic model (Pitz et al. 2007) were found. Thus, the objectives of this work are twofold. The first objective is to collect new autoignition data in an RCM at a higher pressure of 50 bar that include LTC range

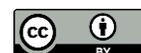

(690–910 K). The datasets include the pressure history that relates to the heat release rates in the RCM, the first stage ignition delay time that is characteristic of the low temperature combustion, and the total ignition delay time that corresponds to hot ignition in the engine. The second objective of this paper is to update the reaction pathways and rate constants of important reactions in the kinetic model of Pitz et al. (2007) and use the previously and newly obtained RCM data to validate the updated LLNL model.

## 2. Methods

2.1 Experimental Methods

The experimental facility consists of a rapid compression machine, a fuel mixture preparation facility, and diagnostics. For mixture preparation, the fuel and oxidizer pre-mixtures are prepared in a stainless steel mixing tank. The volume of the tank is approximately 17 L so that many experiments can be run from a single batch. The liquid fuel (methyl-cyclohexane, 99.0% purity) is massed to a precision of 0.01 g in a syringe before being injected into the mixing tank through a septum. The proportions of oxygen (99.9999% purity), nitrogen (99.9995% purity), and argon (99.9999% purity) are determined by specifying the oxidizer composition, the equivalence ratio, and the total mass of fuel. The gases are added to the mixing tank manometrically at room temperature. The mixture is stirred by a magnetic vane. The mixing tank, reaction chamber, and all lines connecting them are equipped with heaters to prevent condensation of the fuel. After filling the tank, the heaters are turned on and the system is allowed approximately 1.5 hours to equilibrate. This procedure has been validated previously in studies by Weber et al. (2011), Das et al. (2012), and Kumar et al. (2009). In these studies, the concentration of n-butanol, n-decane, and water were verified by GCMS, GC-FID, and GC-TCD, respectively.

Three different mixtures of MCH/$O_2$/$N_2$/Ar are prepared in this study, as outlined in Table 1. These mixtures (denoted as Mix #1–3) match the mixtures prepared in our previous work on MCH in the RCM (Mittal et al. 2009). The equivalence ratios corresponding to Mix #1–3 are 1.0, 0.5, and 1.5, respectively. As in the previous RCM experiments, the mole fraction of MCH is held constant and the mole fraction of $O_2$ is varied to adjust the equivalence ratio. In addition, the relative proportions of $O_2$, $N_2$, and Ar are adjusted so that the same specific heat ratio is maintained in the three mixtures. The approach of maintaining constant specific heat ratio means that the pressure and temperature at the end of compression will be identical for experiments using the same initial pressure, initial temperature, and compression ratio across all three equivalence ratios.

**Table 1: Molar Proportions of Reactants**

| Mix # | $\phi$ | MCH | $O_2$ | $N_2$ | Ar |
|---|---|---|---|---|---|
| 1 | 1.0 | 1 | 10.5 | 12.25 | 71.75 |
| 2 | 0.5 | 1 | 21.0 | 0.00 | 73.50 |
| 3 | 1.5 | 1 | 7.0 | 16.35 | 71.15 |

The RCM used for these experiments is a pneumatically-driven/hydraulically-stopped arrangement. At the start of an experimental run the piston rod is held in the retracted position by hydraulic pressure while the reaction chamber is vacuumed to less than 1 Torr. Then, the reaction chamber is filled with the required initial pressure of test gas mixture from the mixing tank. The compression is triggered by releasing the hydraulic pressure. The piston assembly is driven forward to compress the test mixture by high pressure nitrogen. The gases in the test section are brought to the compressed pressure ($P_C$) and compressed temperature ($T_C$) conditions in approximately 30–50 milliseconds. The piston in the reaction chamber is machined with specifically designed crevices to ensure that the roll-up vortex effect is suppressed and homogeneous conditions in the reaction chamber are promoted. In the present operation procedure $P_C$ and $T_C$ can be varied independently by adjusting the Top Dead Center (TDC) piston clearance, the stroke of the piston, the initial temperature ($T_0$), and the initial pressure ($P_0$) of the test charge. The pressure in the reaction chamber is monitored during and after compression by a Kistler Type 6125B dynamic pressure transducer. During the filling of the mixing tank and reaction chamber prior to compression, the pressure is monitored by an Omega Engineering PX-303 static pressure transducer.

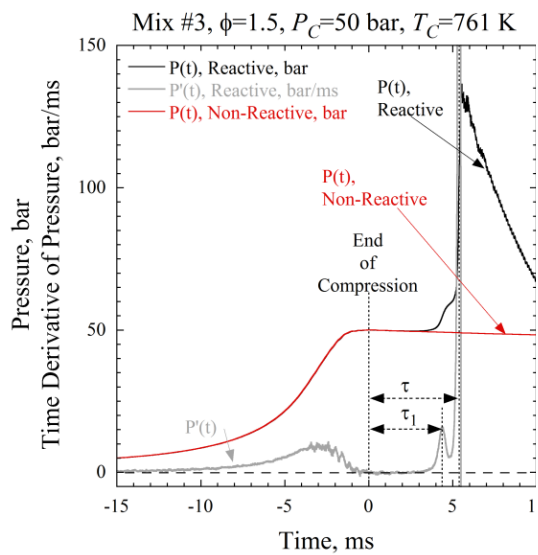

**Figure 1: Representative pressure trace indicating the definition of the ignition delays and the corresponding non-reactive pressure trace.**

Figure 1 shows a representative pressure trace from these experiments at $P_C = 50$ bar, $T_C = 761$ K, and $\phi = 1.5$. The definitions of the end of compression and the ignition delays are indicated on the figure. The end of compression time is defined as the time when the pressure reaches its maximum before first stage ignition occurs, or for cases where there is no first stage ignition, the maximum pressure before the overall ignition occurs. The first stage ignition delay is the time from the end of compression until the first peak in the time derivative of the pressure. The overall ignition delay is the time from the end of compression until the largest peak in the time derivative of the pressure.

Due to heat loss from the test mixture to the cold reactor walls, the pressure and temperature will drop after the end of compression. To properly account for this effect in numerical simulations, a non-reactive pressure trace is taken that corresponds to each unique $P_C$ and $T_C$ condition studied. The non-reactive pressure trace is acquired by replacing the oxygen in the oxidizer with nitrogen, so that a similar specific heat ratio is maintained, but the heat release due to exothermic oxidation reactions is eliminated. A representative non-reactive pressure trace is shown in Fig. 1. This non-reactive pressure trace is converted to a volume trace for use in simulations in CHEMKIN-Pro (2011) using the temperature dependent specific heat ratio.

Each unique $P_C$ and $T_C$ condition is repeated at least 6 times to ensure repeatability of the experiments. The experiment closest to the mean of the runs at a particular condition is chosen for analysis and presentation. The standard deviation of all of the runs at a condition is less than 10% of the mean in all cases. Furthermore, to ensure reproducibility, each new mixture preparation is checked against a previous experiment.

2.2  Numerical Methods

Simulations are performed using the Closed Batch Homogeneous Reactor model in CHEMKIN-Pro. The reactor volume is prescribed as a function of time by the non-reactive volume trace described earlier. This type of simulation captures the heat loss effects during the compression stroke and the post-compression event, which allows the simulation to more accurately describe the actual thermodynamic conditions in the reaction chamber. It also captures the effect of any reactions that occur during compression. This type of simulation is referred to as a VPRO simulation.

VPRO simulations are used to calculate the temperature at the end of compression, $T_C$. This temperature is used as the reference temperature for reporting the ignition delay. This approach requires the assumption of an adiabatic core of gases in the reaction chamber, which is facilitated on the present RCM by the creviced piston described previously. Simulations to determine $T_C$ are conducted with and without reactions in the reaction mechanism. If there is no significant reactivity (and hence heat release) during the compression stroke, the pressure and temperature at TDC are the same whether or not reactions are included in the simulation.

2.3  Mechanism Development

The LLNL MCH chemical kinetic model has been updated to reflect new chemical kinetic information that became available since the publication of our 2007 mechanism (Pitz et al. 2007). The following submodels in the mechanism have been replaced: the $C_1$-$C_4$ base chemistry with the AramcoMech version 1.3 (Metcalfe et al. 2013); the aromatics base chemistry with the latest LLNL-NUIG model (Nakamura et al. 2013); and the cyclohexane submodel with a more recent version from Silke et al. (2007). Also, there are additional specific updates of the MCH mechanism. The abstraction reactions from MCH are replaced using recent experimentally measured values (Sivaramakrishnan et al. 2009) and standardized using the latest LLNL reaction rate rules (Sarathy et al. 2011). The previous 2007 MCH model lumped many of unsaturated ring products of MCH. These species have now been expanded to include all the relevant isomers with their associated reaction paths and rate constants. The model now tracks the intermediate unsaturated methyl-cyclohexane species with much more fidelity and predicts their experimentally measured concentrations with much more accuracy (Pitz et al. 2013).

Regarding the low temperature chemistry portion of the chemical kinetic mechanism, there have been many updates. For the R+$O_2$ reactions involving the cyclohexane ring in MCH, the *ab initio* rate constants computed by Fernandes et al. (2009) were used. They computed the rate constants from 1 to 50 bar over a temperature range of 500–900 K. Since $RO_2$ isomerization rate constants were available for cyclohexane but not for the case of a methyl substitution on the ring, *ab*

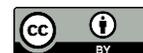

*initio* calculations were performed at LLNL by the authors for the case of a six membered ring where the –OO group is attached at the beta position relative to the methyl group. This reaction was shown to potentially have a large rate constant because of its low activation energy as calculated by Yang et al. (2010). Because we needed the pre-exponential factor that was not reported in the work by Yang et al. (2010), we calculated the high pressure rate constant using the CBS-QB3 method including tunneling (Montgomery et al. 2000). The reactant, product, and transition state geometries and frequencies were calculated using the Gaussian09 suite of programs (Frisch et al. 2009). The lowest energy conformations were obtained from relaxed scans around each rotor in 60° increments using B3LYP/6-31+G(d,p). Reaction rates and Eckart tunneling factors were determined using ChemRate (Mokrushin et al. 2009). The transition state was confirmed by the presence of a single imaginary frequency that corresponds to the reaction pathway, as well as IRC calculations using B3LYP/6-31+G(d,p). The reaction rate was determined using a rigid rotor harmonic oscillator approximation with corrections for hindered rotors. Rotational barriers were determined via relaxed scans in 10° increments using B3LYP/6-31+G(d,p). Our computed rate constant has an activation energy which is 2 kcal higher than the value of Yang et al. (2010). We attribute this difference to the choice of the conformer for the reactant $RO_2$ species. The lowest lying conformer is normally chosen for the reactant. We found a lower energy conformer than Yang et al. (2010) chose in their study.

Rate constants for R+$O_2$ involving a tertiary site were not available from Fernandes et al. (2009), so these rate constants were estimated. Since pressure dependent rate constants were not available for all the R+$O_2$ reactions for MCH, we used the high pressure rate constants for all of them for consistency. For the rate constants obtained from Fernandes et al. (2009) where the high pressure rate constant was not reported, we used the 50 bar rate constant. Since the pressures of interest in this study are 15 bar and above, we consider this a good assumption. We also consider this a good assumption for pressures encountered in internal combustion engines.

For the thermodynamic parameters for species, we have adapted those in AramcoMech (Metcalfe et al. 2013) for $C_1$-$C_4$ species. For other new species in the mechanism such as the unsaturated cyclic species derived from MCH, THERM (Ritter et al. 1991) was used to calculate the thermodynamic parameters.

## 3. Results and Discussion

The experimental ignition delays measured at the three equivalence ratios and compressed pressure of 50 bar are shown in Figure 2. The open symbols are the overall ignition delay and the filled symbols are the first stage ignition delays. The vertical error bars on the experimental data represent twice the standard deviation of all of the experiments at that condition. The temperature uncertainty has been estimated previously as no greater than 3.5 K by Mittal et al. (2009).

The negative temperature coefficient (NTC) region is an important feature of low temperature ignition where the ignition delay time decreases with increasing temperature. The NTC region of the overall ignition delay is evident in Fig. 2 for the $\phi = 1.5$ case and approximately includes the temperature range 775–840K. For $\phi = 1.5$, first stage ignition is evident for conditions in the range of $T_C = 740$–800K.

For $\phi = 1.0$, the NTC region of the overall ignition delay could not be completely resolved. Only three conditions in the low temperature region and three conditions in the high temperature region are shown in Fig. 2. The experimental pressure traces during the compression stroke for intermediate temperature conditions were seen to deviate from their non-reactive counterparts, demonstrating appreciable reactivity therein. Hence, those data are not included in Fig. 2.

For the experiments at $\phi = 0.5$, only three data points in the low temperature region are reported and none of them exhibits two-stage ignition. As temperature is increased further, noticeable reactivity during the compression stroke is evident.

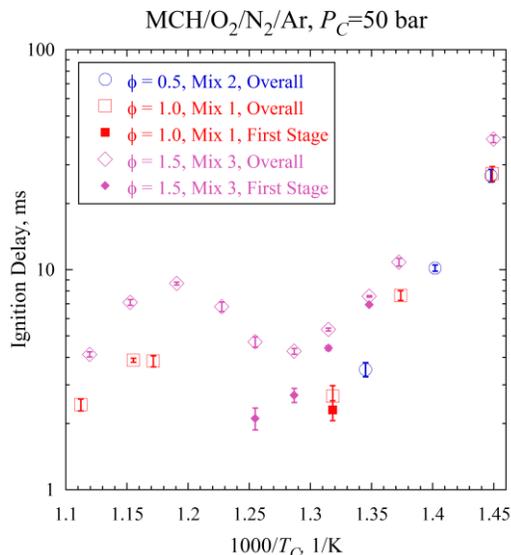

**Figure 2: Experimental ignition delays measured at $P_C = 50$ bar for the mixture conditions in Table 1**

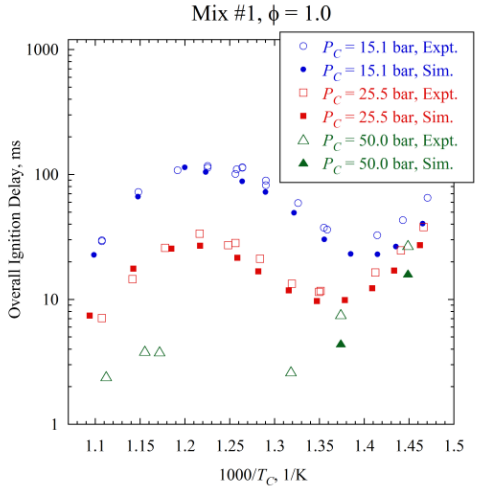

**Figure 3: Comparison of experimental and simulated overall ignition delays of Mix #1 for three pressures.**

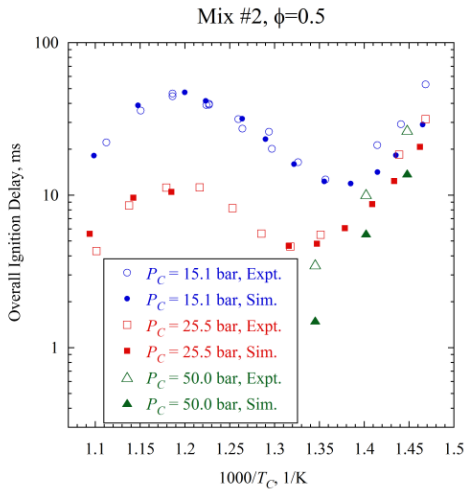

**Figure 4: Comparison of experimental and simulated overall ignition delays of Mix #2 for three pressures.**

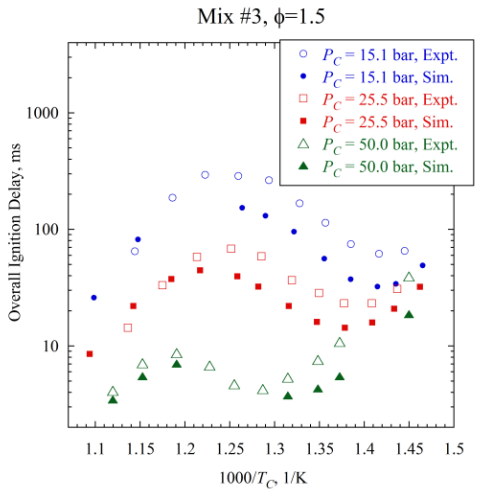

**Figure 5: Comparison of experimental and simulated overall ignition delays of Mix #3 for three pressures**

As stated earlier, the mole fraction of the fuel is held constant in this study, while the mole fraction of the oxygen is changed to modify the equivalence ratio. Figure 2 demonstrates that the $\phi = 0.5$ case is the most reactive (as judged by the inverse of the ignition delay), and the $\phi = 1.5$ case is the least reactive. As has been shown for other fuels, including butanol (Weber et al. 2011) and Jet-A (Kumar et al. 2010), decreasing the equivalence ratio by increasing the oxygen mole fraction but holding the fuel mole fraction constant increases the reactivity. Furthermore, it can be seen in Fig. 2 that the beginning of the NTC region appears to shift to higher temperature as equivalence ratio decreases. This also matches the trend seen in the previous work for MCH in the RCM (Mittal et al. 2009).

A comparison of the experimentally measured overall ignition delays (open symbols) and the model of the current version (closed symbols), which will be referred to as v9f hereafter, is shown in Figures 3-5. In addition, a comparison of the experimentally measured first stage ignition delays (open symbols) and the v9f model (closed symbols) are shown in Figures 6-8. The experiments include the new work being presented here at $P_C = 50$ bar in addition to the previous RCM experiments at $P_C = 15.1$ and $25.5$ bar. The simulations are the VPRO type of simulations. For some computational cases, substantial heat release during the compression stroke caused the computed pressure to depart from the non-reactive profile. Therefore, these cases are not shown in Figures 3-8.

Comparing the predicted and measured overall ignition delay times, the simulations reasonably well predict the ignition delay ($\pm 50\%$) for the lean and stoichiometric cases. For lower temperatures at these two equivalence ratios, the experimental ignition delays are under predicted by the model. For the rich case, the simulations under-predict the ignition delay but the results improve as pressure increases.

The first stage ignition delays for all of the pressure and equivalence ratios are under predicted, but are within 50% of the experimental values. Furthermore, for all of the equivalence ratios tested at $P_C = 50$ bar, it is also of interest to note that while the experimental pressure traces exhibit single-stage ignition, the simulated results show two-stage ignition response. Nevertheless, the present mechanism update is a marked improvement from the comparison performed by Mittal et al. (2009) who found that the ignition delays were uniformly over-predicted by the previous LLNL mechanism by Pitz et al. (2007).

## 4. Conclusions

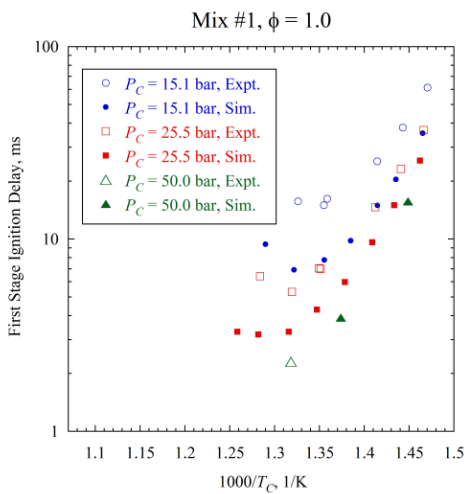

**Figure 6: Comparison of experimental and simulated first stage ignition delays of Mix #1 for three pressures.**

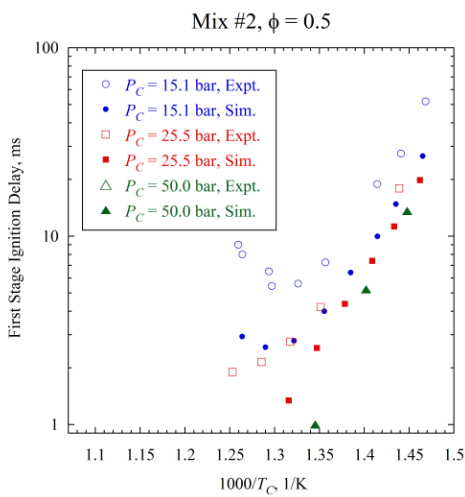

**Figure 7: Comparison of experimental and simulated first stage ignition delays of Mix #2 for three pressures.**

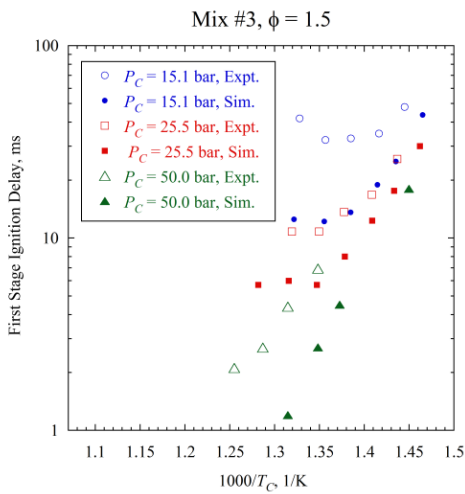

**Figure 8: Comparison of experimental and simulated first stage ignition delays of Mix #3 for three pressures.**

In this work, newly measured experimental ignition delays of methyl-cyclohexane mixtures have been presented. For three mixtures of MCH/$O_2$/$N_2$/Ar, experiments were conducted at compressed pressure of 50 bar and over the compressed temperature range 690-910 K. The experimental data measured in this study are further compared to the data previously collected in the RCM at 15.1 and 25.5 bar. The new data shows similar trends in reactivity as the previous data. Furthermore, an updated model of MCH combustion has been developed. Many updates have been made to the reactions and pathways in the model, especially including the low-temperature pathways. Due to these updates, the ability of the model to correctly predict the ignition delays of MCH is substantially improved. Comparison of the new and existing data with the newly updated model shows that ignition delays for most conditions are predicted to within ±50% by the new model. Although these improvements are encouraging, further work is still needed, e.g. on improving predictions of the first stage ignition delays.

**Acknowledgements**

The work at the University of Connecticut was supported as part of the Combustion Energy Frontier Research Center, an Energy Frontier Research Center funded by the US Department of Energy, Office of Science, Office of Basic Energy Sciences, under Award Number DE-SC0001198 and by the National Science Foundation under Grant No. 0932559. This work at LLNL was supported by U.S. Department of Energy, Office of Vehicle Technologies, and performed under the auspices of the U.S. Department of Energy by Lawrence Livermore National Laboratory under Contract No. DE-AC52-07NA27344.

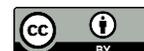